\documentclass[reprint,superscriptaddress,nofootinbib,amsmath,amssymb,physrev]{revtex4-2}

\pdfoutput=1
\usepackage{float}
\usepackage{xcolor}
\usepackage{graphicx}
\usepackage{dcolumn}
\usepackage{bm,lipsum}
\usepackage{siunitx}
\usepackage[colorlinks=true, linkcolor=blue, citecolor=blue, urlcolor=blue]{hyperref}

\newcommand{\ZIB}{Zuse Institute Berlin, 14195 Berlin, Germany}
\newcommand{\JCM}{JCMwave GmbH, 14050 Berlin, Germany}
\newcommand{\TUB}{Institut für Physik und Astronomie, Technische Universität Berlin, 10623 Berlin, Germany}
\newcommand{\DTU}{Department of Electrical and Photonics Engineering, Technical University of Denmark, 2800 Kongens Lyngby, Denmark}

\usepackage{amsmath}
\usepackage{graphicx}
\usepackage{booktabs}
\usepackage{tabularx}
\usepackage{xcolor,colortbl}
\usepackage[T1]{fontenc}

\begin{document}

\title{High Purcell enhancement in all-TMDC nanobeam resonator designs with \\ active monolayers for nanolasers}
\author{Felix Binkowski}
\thanks{These authors contributed equally to this work.}
\affiliation{\JCM}
\affiliation{\ZIB}
\author{Aris~Koulas-Simos}
\thanks{These authors contributed equally to this work.}
\affiliation{\TUB}
\author{Fridtjof~Betz}
\affiliation{\ZIB}
\author{Matthias~Plock}
\affiliation{\JCM}
\author{Ivan~Sekulic}
\affiliation{\JCM}
\author{Phillip~Manley}
\affiliation{\JCM}
\author{Martin~Hammerschmidt}
\affiliation{\JCM}
\author{Philipp-Immanuel~Schneider}
\affiliation{\JCM}
\author{Lin~Zschiedrich}
\affiliation{\JCM}
\author{Battulga~Munkhbat}
\affiliation{\DTU}
\author{Stephan~Reitzenstein}
\affiliation{\TUB}
\author{Sven~Burger}
\affiliation{\JCM}
\affiliation{\ZIB}

\begin{abstract}
We propose a nanobeam resonator incorporating an active monolayer, designed to achieve a high Purcell enhancement. The resonator is fully composed of transition-metal-dichalcogenide materials and intended to operate as a high-$\beta$-factor nanolaser. A theoretical framework that models and optimizes the Purcell enhancement associated with the emission from atomically thin layers is developed. This framework is based on a resonance expansion, enabling spectral resolution of physical quantities governed by high-$Q$ resonances. The numerical optimization of the resonator leads to the presence of a high-$Q$ resonance supporting a strong electric field confinement in the monolayer to maximize the modal gain.
\end{abstract}

\maketitle
\section{Introduction}
Two-dimensional (2D) materials, such as atomically thin layers of transition-metal-dichalcogenides (TMDCs)~\cite{Mak_2010,Splendiani_2010,Wang_2018}, graphene~\cite{Novoselov_2004,Bonaccorso_2010}, or hexagonal boron nitride~\cite{Dean_2010,Tran_2016}, emerge as promising platforms in the field of nanophotonics due to their ability to tailor light-matter interactions. The incorporation of 2D materials into nanophotonic resonators enables, e.g., the control of their emission properties~\cite{Sortino_2019} and applications in nonlinear photonics~\cite{Seyler_2015,Autere_2018} and quantum technology~\cite{Tonndorf_2015,Tran_2019}.

\footnote{This work has been published:\\
F. Binkowski et al., Phys. Rev. B \textbf{112}, 235410 (2025).\\
DOI: \href{https://doi.org/10.1103/nxh9-dhvx}{10.1103/nxh9-dhvx}}

Nanolasers are ultra-compact coherent light sources that combine nanoscale optical resonators with low-dimensional gain materials~\cite{Hill_2014,Deng_2021}. By leveraging strong light confinement, they offer potential applications in, e.g., integrated photonics~\cite{Munnelly_2017}, quantum communication, and sensing~\cite{Oulton_2009}. A key figure in nanolaser performance is the Purcell effect, which describes the modified spontaneous emission rate of an emitter due to the presence of a resonator~\cite{Purcell_1946}. A high Purcell enhancement lowers the lasing threshold and facilitates efficient light–matter interaction in the weak-coupling regime of cavity quantum electrodynamics~\cite{Purcell_1946, Wu_Nature_2015}. The Purcell enhancement scales with increasing light confinement, which is reflected by a low mode volume~\cite{Wu_ACSPhot_2021}. In particular, small mode volumes not only boost the Purcell enhancement, but also enhance the fraction of spontaneous emission funneled into the lasing mode, quantified by the $\beta$-factor. A high Purcell enhancement in such resonators can thus drive the $\beta$-factor close to unity, which is essential for achieving low-threshold or thresholdless lasing~\cite{Khajavikhan_2012}. The Purcell enhancement also scales with increasing resonator quality ($Q$) factor~\cite{Wu_ACSPhot_2021}. While a high $Q$-factor is generally desirable for a high Purcell enhancement, excessively high $Q$-factors can result in a very narrow resonance linewidth, which reduces the spectral overlap with the gain medium and thereby limits the coupling efficiency in systems with realistic inhomogeneously broadened emission spectra~\cite{Lodahl_2015}.

Traditionally, nanolasers combine dielectric or metallic cavities with III–V semiconductor gain media~\cite{Reitzenstein_2006,Jagsch_2018,Koulas-Simos_2022,Koulas-Simos_LPOR_2022}. However, monolayer TMDCs have recently emerged as compelling alternatives due to their strong excitonic response, large excitonic binding energies, and direct bandgaps in the monolayer limit. These properties have enabled numerous experimental demonstrations of TMDC-based nanolasers over the past decade~\cite{Wu_Nature_2015,Ye_2015,Salehzadeh_2015,Li_2017,Koulas-Simos_2024}, positioning them as viable 2D~gain materials alongside III–V quantum wells. Beyond their role as gain media, TMDCs are recently also being explored as resonator materials~\cite{Munkhbat_2023,Alekseev_2025}. Their high refractive index supports strong light confinement and high-$Q$ resonances, while their layered nature enables mechanical stacking and seamless integration into planar photonic platforms~\cite{Geim_2013,Palekar_2024}. This dual functionality is driving interest in fully TMDC-based nanolasers, opening avenues for compact, scalable, and all-TMDC photonic systems.

In the emission process of a 2D material, multiple emitters distributed across the 2D material contribute through spatially extended excitation and collective emission processes~\cite{Plankensteiner_2019}. In the case of TMDC monolayers, the excitions are spatially delocalized over finite coherence lengths~\cite{Chernikov_2015}. However, for the purpose of comparing different resonator designs, it can be sufficient to consider a single emitter placed at the position of maximum electric field intensity. This is especially valid when the resonators have similar geometries, such that the spatial profiles of their high-$Q$ resonance modes are comparable. While the inclusion of many emitters can lead to a higher total Purcell enhancement~\cite{Carminati_2022}, the relative differences between resonator designs can be captured by considering the emission from a single emitter. The Purcell enhancement of an emitter can be calculated by solving Maxwell's equations with a dipole-like current density as the source term. Advanced numerical methods must be used to accurately calculate the scattered electric field caused by the singular source~\cite{Awada_IEEE_1997}. An alternative approach is a resonance expansion, where the scattered electric field is decomposed into a weighted sum of resonance modes~\cite{Lalanne_QNMReview_2018,Nicolet_2023}. These resonance modes are the solutions to the source-free Maxwell's equations, whose numerical treatment is accompanied by conceptual differences compared to the treatment of a source term, e.g., no singular electric field has to be considered in the numerical implementation~\cite{Lalanne_QNM_Benchmark_2018,Demesy_ComputPhysComm_2020}. One advantage of a resonance expansion is that, when calculating the Purcell enhancement of systems with high-$Q$ resonances, the resulting sharp spectral peak of the Purcell enhancement can be efficiently resolved.

\begin{figure}
\includegraphics[width=0.49\textwidth]{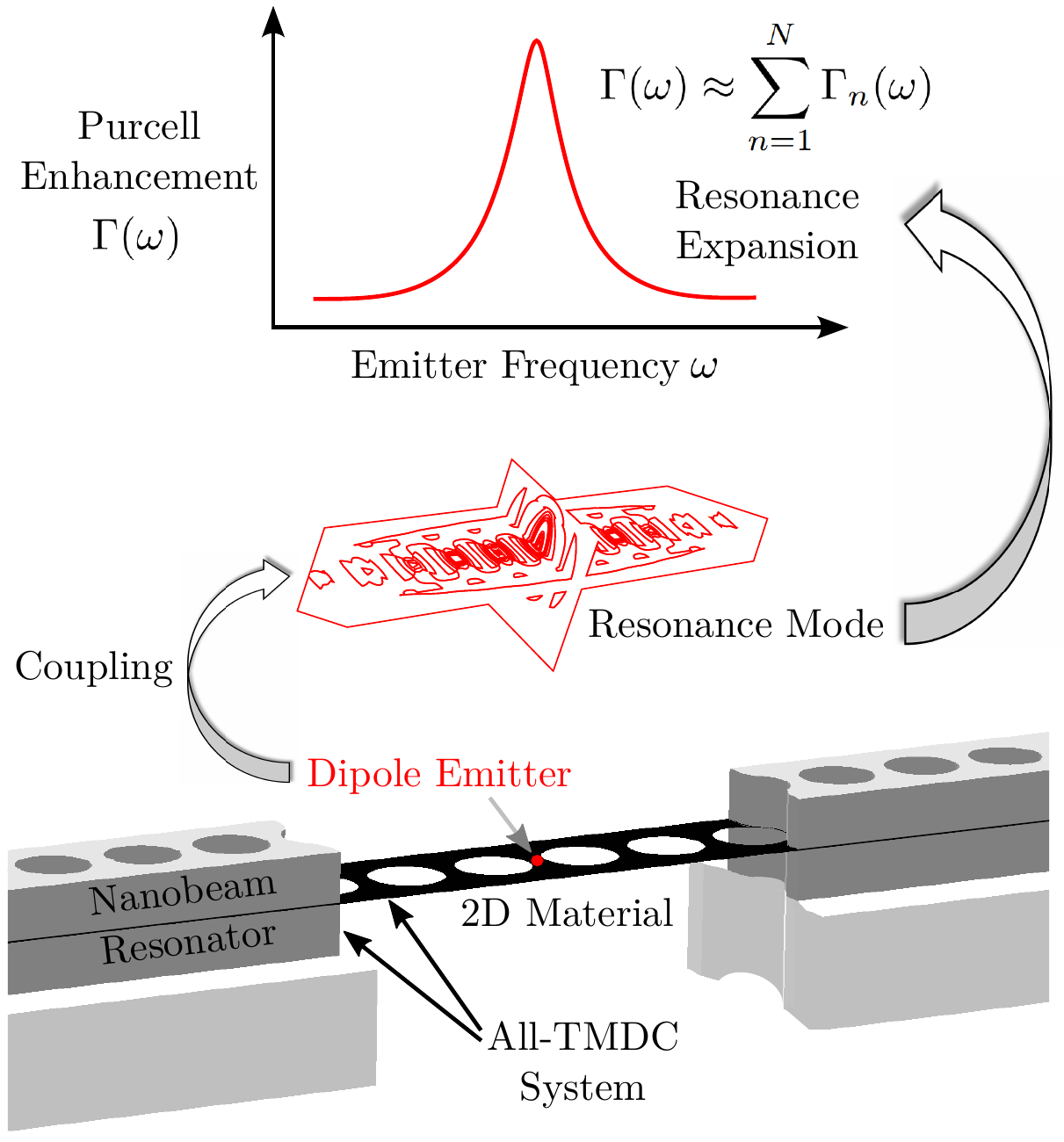}
\caption{\label{fig1}
Purcell enhancement in all-TMDC nanobeam resonators. The 2D material, which is a TMDC monolayer, acts as an active layer. In the central region of the sketched resonator, the materials of the resonator are omitted for illustration purposes. The emission process can be modeled by a dipole emitter that couples with the resonance modes of the system. The total Purcell enhancement $\Gamma(\omega)$, where $\omega$ is the emission frequency of the dipole emitter, can be calculated by a resonance expansion based on individual modal contributions $\Gamma_n(\omega)$.}
\end{figure}

In this work, we propose a nanobeam resonator for nanolasing applications, consisting of an active 2D material in the form of a TMDC monolayer embedded in an optical waveguide with periodically arranged holes, which is also composed of a TMDC material. We develop a theoretical and computational framework to model and optimize the Purcell enhancement of such systems. The emission process of the monolayer is modeled by a single dipole emitter at the center of the nanobeam resonator. The optimization is based on resonance expansion and a constraint on the $Q$-factor is imposed to ensure that the resulting spectral peak of the Purcell enhancement is sufficiently broad for real-world scenarios. The optimized nanobeam resonator exhibits a resonance with a $Q$-factor of $9.92 \times 10^4$ and a strong confinement of the electric field in the monolayer. The optimization results in a Purcell enhancement of $4.6\times10^3$.

\section{Calculation of the Purcell enhancement}
In nanophotonics, in the steady-state regime, light scattering by an open material system can be investigated by solving the time-harmonic Maxwell's equation equipped with open boundary conditions,
\begin{align}
	\nabla \times  \mu_0^{-1} 
	\nabla \times \mathbf{E}(\mathbf{r},\omega)  -
	\omega^2\epsilon(\mathbf{r},\omega) \mathbf{E}(\mathbf{r},\omega)  = 
	i\omega\mathbf{J}(\mathbf{r}), \label{eq:maxwell}
\end{align}
where $\mathbf{E}(\mathbf{r},\omega) \in \mathbb{C}^3$ is the scattered electric field, $\mathbf{J}(\mathbf{r}) \in \mathbb{C}^3$ is an optical source term, $\omega \in \mathbb{C}$ is the angular frequency, and $\mathbf{r} \in \mathbb{R}^3$ is the considered spatial position. The spatial distribution of material and the material dispersion are described by the permittivity $\epsilon(\mathbf{r},\omega) = \epsilon_\mathrm{r}(\mathbf{r},\omega) \epsilon_0$, where $\epsilon_\mathrm{r}(\mathbf{r},\omega)$ is the relative permittivity and $\epsilon_0$ is the vacuum permittivity. When optical frequencies are taken into account, the permeability $\mu(\mathbf{r},\omega) = \mu_\mathrm{r}(\mathbf{r},\omega) \mu_0$ typically equals the vacuum permeability $\mu_0$.

To investigate the Purcell enhancement of a point-like source, we model the source term by the current density of a dipole emitter, $\mathbf{J}(\mathbf{r}) = \mathbf{j} \delta\left(\mathbf{r}-\mathbf{r}^\prime\right)$, where $\mathbf{r}^\prime$ is the location of the dipole emitter, $\delta(\mathbf{r}-\mathbf{r}^\prime)$ is the Dirac delta distribution, and $\mathbf{j} = -i \omega \mathbf{p}$ is the dipole strength vector with the dipole moment $\mathbf{p}$. With this, the Purcell enhancement of the dipole emitter is given by
\begin{align}
    \Gamma(\omega) = - \frac{1}{2} \mathrm{Re}\left(\mathbf{E}(\mathbf{r}^\prime,\omega)\cdot 
    \mathbf{j}^*\right)/\Gamma_\mathrm{b}, \label{eq:purcell}
\end{align}
where $\Gamma_\mathrm{b}$ is the decay rate of the dipole emitter in 
homogeneous background material~\cite{Sauvan_QNMexpansionPurcell_2013,Betz_2022}.

\subsection{Scattering problem}
The spatial singularity of the dipole emitter results in a poor regularity of the resulting electric field $\mathbf{E}(\mathbf{r},\omega)$, which is the solution of the scattering problem given by Eq.~\eqref{eq:maxwell}. This leads to an insufficient numerical convergence of the solution and of derived quantities such as the Purcell enhancement with respect to the numerical discretization parameters.

The subtraction approach~\cite{Awada_IEEE_1997,Zschiedrich_OLED_2013} can be applied to overcome this issue. This approach is based on the availability of an analytically known singular solution $\mathbf{E}_\mathrm{s}(\mathbf{r},\omega)$. For a dipole emitter in homogeneous background material, $\mathbf{E}_\mathrm{s}(\mathbf{r},\omega)$ is the homogeneous Green's function. Quasi-analytical solutions are also available when the source is located at or close to a planar material interface~\cite{Paulus_PRE_2000}. 

In the following, the subtraction approach is derived. For simplification, we omit the spatial and frequency dependencies of the electric field and write $\mathbf{E}$ instead of $\mathbf{E}(\mathbf{r},\omega)$. Considering homogeneous background material in the vicinity of the dipole emitter yields
\begin{align}
	\nabla \times \mu_0^{-1} 
	\nabla \times \mathbf{E}_\mathrm{s} -
	\omega^2\epsilon_\mathrm{b} \mathbf{E}_\mathrm{s} =
	i\omega\mathbf{j}\delta(\mathbf{r}-\mathbf{r}^\prime), \nonumber
\end{align}
where $\epsilon_\mathrm{b}$ is the constant permittivity of the background material, i.e., the material in which the dipole emitter is located. The electric field $\mathbf{E}$, which is a solution to the original problem given by Eq.~\eqref{eq:maxwell}, can be written as $\mathbf{E} = \mathbf{E}_\mathrm{s} + \mathbf{E}_\mathrm{c}$, where $\mathbf{E}_\mathrm{c}$ is a correction field. Inserting this into Eq.~\eqref{eq:maxwell} gives
\begin{align}
	&\nabla  \times  \mu_0^{-1} 
	\nabla  \times  \left(\mathbf{E}_\mathrm{s} 
	+ \mathbf{E}_\mathrm{c}\right)  
	-  \omega^2\epsilon \left(\mathbf{E}_\mathrm{s} 
	+ \mathbf{E}_\mathrm{c}\right)  
	\nonumber \\
	&= \nabla  \times  \mu_0^{-1} 
	\nabla  \times  \mathbf{E}_\mathrm{s}  
	-  \omega^2\epsilon \mathbf{E}_\mathrm{s}  
	+ \nabla  \times  \mu_0^{-1} 
	\nabla  \times  \mathbf{E}_\mathrm{c}  
	-  \omega^2\epsilon \mathbf{E}_\mathrm{c}  
	\nonumber \\
	&=  \nabla  \times  \mu_0^{-1} 
	\nabla  \times  \mathbf{E}_\mathrm{s}  
	-  \omega^2 \epsilon_\mathrm{b} \mathbf{E}_\mathrm{s} 
	-  \omega^2 \left(\epsilon -\epsilon_\mathrm{b}\right) \mathbf{E}_\mathrm{s} 
	\nonumber \\
	& \hspace{1cm} + \nabla  \times  \mu_0^{-1} 
	\nabla  \times  \mathbf{E}_\mathrm{c}  
	-  \omega^2\epsilon \mathbf{E}_\mathrm{c}  
	\nonumber \\
	&= 	i\omega\mathbf{j}\delta(\mathbf{r}-\mathbf{r}^\prime)
	-  \omega^2 \left(\epsilon -\epsilon_\mathrm{b}\right) \mathbf{E}_\mathrm{s} 
	\nonumber \\
	& \hspace{1cm} + \nabla  \times  \mu_0^{-1} 
	\nabla  \times  \mathbf{E}_\mathrm{c}  
	-  \omega^2\epsilon \mathbf{E}_\mathrm{c}  
	\nonumber \\
	&= 	i\omega\mathbf{j}\delta(\mathbf{r}-\mathbf{r}^\prime). \nonumber
\end{align}
Rearranging yields Maxwell's equation for the correction field,
\begin{align}
\nabla  \times  \mu_0^{-1} 	\nabla  \times  \mathbf{E}_\mathrm{c}  
	-  \omega^2\epsilon \mathbf{E}_\mathrm{c}
	= \omega^2 \left(\epsilon -\epsilon_\mathrm{b}\right) \mathbf{E}_\mathrm{s}, \label{eq:Maxwell_correction}
\end{align}
where the right hand side becomes zero in the vicinity of the dipole emitter. Further away from the location of the emitter, the field $\mathbf{E}_\mathrm{s}$ is sufficiently smooth. Therefore, instead of solving Eq.~\eqref{eq:maxwell} to obtain directly $\mathbf{E}$, one can solve Eq.~\eqref{eq:Maxwell_correction}, where the regularity of its solution $\mathbf{E}_\mathrm{c}$ is suitable for a numerical realization. The solution to Eq.~\eqref{eq:maxwell} can then be obtained indirectly by $\mathbf{E} = \mathbf{E}_\mathrm{s} + \mathbf{E}_\mathrm{c}$. This splitting can be used to evaluate the expression for the Purcell enhancement given by Eq.~\eqref{eq:purcell}, where $\mathrm{Re}(\mathbf{E}_\mathrm{s}(\mathbf{r}^\prime,\omega) \cdot \mathbf{j}^*)$ is analytically known and $\mathbf{E}_\mathrm{c}$ is smooth~\cite{Zschiedrich_OLED_2013}.

\begin{figure*}
\includegraphics[width=0.98\textwidth]{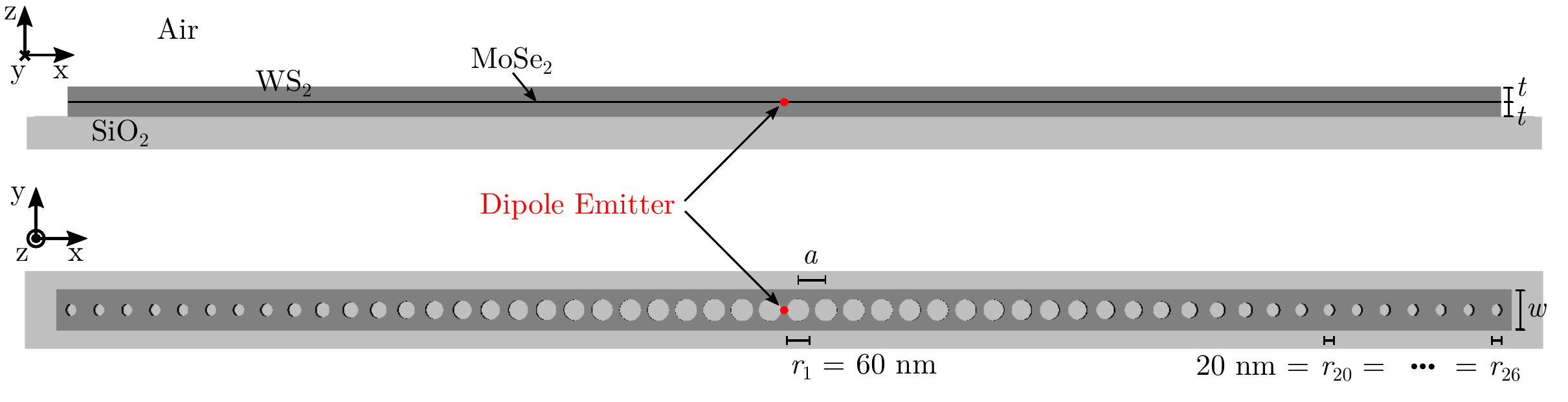}
\caption{\label{fig2}
Sketch of the nanobeam resonator investigated. The 2D material is given by a $1\,\mathrm{nm}$ thick $\mathrm{MoSe}_2$ layer as TMDC monolayer, which lies between two TMDC layers composed of $\mathrm{WS}_2$. The 1D photonic crystal is realized by holes with decreasing radii towards the outer edges of the resonator. The system is placed on an amorphous $\mathrm{SiO}_2$ substrate and it is surrounded by air. The emission of the $\mathrm{MoSe}_2$ layer is modeled by a single dipole emitter located at the center of the resonator. The free parameters of the resonator are given by the lattice constant $a$, the width $w$ of the resonator, and the thickness $t$ of each $\mathrm{WS}_2$ layer.
}
\end{figure*}

\subsection{Resonance problem}
An alternative to the approach of computing the Purcell enhancement based on the scattered electric field $\mathbf{E}(\mathbf{r},\omega)$ from Eq.~\eqref{eq:maxwell} is to solve Eq.~\eqref{eq:maxwell} without a source term, i.e., to solve a resonance problem, and a subsequent expansion of $\mathbf{E}(\mathbf{r},\omega)$ based on the calculated resonance modes $\mathbf{E}_n(\mathbf{r}) \in \mathbb{C}^3$ and the corresponding resonance frequencies $\omega_n \in \mathbb{C}$. For this, the modes are normalized such that $\int_\Omega \left[ {\mathbf{E}_n}(\mathbf{r})\cdot \frac{\partial \omega \epsilon(\mathbf{r},\omega)}{\partial \omega} {\mathbf{E}_n}(\mathbf{r}) - \mu_0 {\mathbf{H}_n}(\mathbf{r})\cdot {\mathbf{H}_n}(\mathbf{r}) \right] dV = 1$, where $\mathbf{H}_n$ is the magnetic field corresponding to the resonance mode and $\Omega$ is the computational domain~\cite{Sauvan_QNMexpansionPurcell_2013,Lalanne_QNMReview_2018}. The term $\mathrm{Re}(\mathbf{E}(\mathbf{r}^\prime,\omega)\cdot \mathbf{j}^*)$ from Eq.~\eqref{eq:purcell} can then be approximated by the expansion~\cite{Betz_2022},
\begin{equation} \label{eq:expansion}
   \mathrm{Re}(\mathbf{E}(\mathbf{r}^\prime,\omega)\cdot \mathbf{j}^*) \approx \mathrm{Re}\left(\sum_{n=1}^{N} \alpha_n(\omega) \mathbf{E}_n(\mathbf{r}^\prime)\cdot \mathbf{j}^*\right),
\end{equation}
where the expansion coefficients $\alpha_n(\omega)$ are given by
\begin{equation} \nonumber
    \alpha_n(\omega) = \frac{i}{\omega_n - \omega}\mathbf{E}_n(\mathbf{r}^\prime) \cdot \mathbf{j}.
\end{equation}
Inserting Eq.~\eqref{eq:expansion} into Eq.~\eqref{eq:purcell} yields
\begin{equation}
\begin{aligned}
    \Gamma(\omega) &\approx - \frac{1}{2} \mathrm{Re}\left(\sum_{n=1}^{N} \alpha_n(\omega) \mathbf{E}_n(\mathbf{r}^\prime)\cdot \mathbf{j}^*\right)/\Gamma_\mathrm{b} \\
    &= \sum_{n=1}^{N}\left( - \frac{1}{2} \mathrm{Re}\left(\alpha_n(\omega) \mathbf{E}_n(\mathbf{r}^\prime)\cdot \mathbf{j}^*\right)/\Gamma_\mathrm{b}\right) \\
    &= \sum_{n=1}^{N} \Gamma_n(\omega), \label{eq:purcell_expansion}
\end{aligned}
\end{equation}
where $\Gamma_n(\omega)$ are the individual modal contributions to the total Purcell enhancement $\Gamma(\omega)$. Assuming that $N$ modes yield a sufficient approximation in Eq.~\eqref{eq:expansion}, this formula can be used for the computation of the Purcell enhancement corresponding to a dipole emitter.  This scenario is sketched in Fig.~\ref{fig1}.

The advantages of solving Eq.~\eqref{eq:maxwell} without a source term and a subsequent resonance expansion of the Purcell enhancement include the following aspects: i) No electric field with a spatial singularity needs to be considered. With very thin layers or in the vicinity of curved material interfaces, the implementation of the subtraction approach can also reach its limits. In contrast, the resonance modes are sufficiently smooth and numerical convergence can be guaranteed. ii) Characteristics such as the $Q$-factor, which is defined by $Q = \mathrm{Re}(\omega_n)/(-2\mathrm{Im}(\omega_n))$, or the field patterns of the resonance modes can be extracted and investigated directly. iii) For expanding the Purcell enhancement based on different source terms, e.g., sources with different spatial positions or polarizations, only one set of resonance modes and frequencies needs to be calculated. This can reduce the computational effort. iv) If fine frequency sampling is required, which is the case when resonances with high $Q$-factors lead to a sharp spectral peak of the Purcell enhancement, then the evaluation of the resonance expansion for the different frequencies causes a negligible computational effort. This point is the main reason why a resonance expansion is employed in the following to calculate the Purcell enhancement.

\section{Nanobeam resonator with an active monolayer}
We focus on a nanobeam resonator design, reflecting its role as a canonical and technologically relevant platform in integrated photonics. Nanobeam resonators naturally combine high-$Q$ resonances, small mode volumes, and waveguiding functionality~\cite{Lalanne_2003,Sauvan_2005}, enabling seamless integration into photonic circuits~\cite{Pernice_2012} and efficient coupling to atomically thin materials~\cite{Li_2017}. Importantly, the physical mechanisms investigated here, Purcell enhancement, resonator–exciton coupling, and all-TMDC integration, are not unique to nanobeam resonators. These concepts can be extended to other resonator geometries, including photonic crystal cavities, microrings, slot waveguides, and hybrid dielectric–plasmonic platforms~\cite{Sriram_2020,Munkhbat_2023}.

We study a nanobeam resonator incorporating a 2D material. The resonator consists of a $1\,\mathrm{nm}$ thick molybdenum diselenide ($\mathrm{MoSe}_2$) monolayer placed between two tungsten disulfide $(\mathrm{WS}_2)$ layers, each with a thickness of $t$. The resonator fully based on TMDC materials has the width $w$, lies on an amorphous silicon dioxide ($\mathrm{SiO}_2$) substrate, and is surrounded by air. The system is designed as a one-dimensional (1D) photonic crystal, where $26$ cylindrical air holes are placed in each of the two mirror symmetry planes. The centers of the holes are separated by the distance $a$, which is the so-called lattice constant of the photonic crystal. The radii $r_k$ of the holes, where $k=1,...,20$, are descending from the center to the outer edges of the resonator. This relation is given by $r_{k} = (r_1-r_{20})\times ((20-k)/19)^2 + r_{20}$, where $r_1 = 60\,\mathrm{nm}$ is the radius of the hole closest to the center of the resonator and $r_{20} = 20\,\mathrm{nm}$ is the radius of the outer hole. Additional six holes with constant radii $r_{21}=...=r_{26} = r_{20}$ are added. This combination of descending and constant radii provides a Gaussian attenuation of the electrical field intensity towards the outer regions of the nanobeam resonator with the highest intensity located at the center of the resonator. The system is sketched in Fig.~\ref{fig2}.

The $\mathrm{MoSe}_2$ monolayer is considered as a gain material for lasing operation. The fundamental principle of lasing relies on the efficient coupling of the light, which is emitted through recombination of charge carriers in the gain material, into the resonance modes of the resonator \cite{Loudon2000}. This process requires strong spatial overlap between the emitting gain material and the electric field distribution of the resonance modes, as well as strong spectral overlap between the emission spectrum of the gain material and the resonance frequencies~\cite{Romeira2018}, both of which are closely linked to high Purcell enhancement. Typically, the emission of the 2D exciton of a $\mathrm{MoSe}_2$ monolayer is expected to change from roughly $\lambda = 790\,\mathrm{nm}$ at room temperature to around $\lambda=750\,\mathrm{nm}$ at cryogenic temperatures \cite{Tongay2012,Tonndorf2013,Sial2018}. In the following, to model the 2D exciton emission, we consider a single circularly polarized dipole emitter, i.e., $\mathbf{j} = [1,1i,0]^T$, emitting at $\lambda=780\,\mathrm{nm}$ and placed at the center of the $\mathrm{MoSe}_2$ monolayer. We set $\epsilon_\mathrm{r} = 27.091$ for $\mathrm{MoSe}_2$~\cite{Hsu2019} and $\epsilon_\mathrm{r} = 2.1346+0039i$ for $\mathrm{SiO}_2$~\cite{RodrguezdeMarcos2016}. For $\mathrm{WS}_2$, we choose an anisotropic material model, given by the permittivity tensor with the diagonal entries $\epsilon_{\mathrm{r},xx} = 17.36$, $\epsilon_{\mathrm{r},yy} = 17.36$, and $\epsilon_{\mathrm{r},zz} = 6.33$~\cite{Munkhbat2022}. The losses of the $\mathrm{WS}_2$ layers, i.e., the imaginary parts of the permittivity tensor, are negligible at $\lambda = 780\,\mathrm{nm}$. Furthermore, for the sake of simplicity, no material dispersion is taken into account. However, the inclusion of losses within the $\mathrm{WS}_2$ layers and material dispersion is possible for the framework presented. Note that, in line with commonly adopted notation, wavelengths $\lambda$ rather than frequencies $\omega$ are used, where $\lambda = 2 \pi c / \omega$.

We stress that the dipole emitter serves as a proxy for the delocalized excitonic response of the MoSe$_2$ monolayer. The local Purcell enhancement evaluated at the position of maximum electric field intensity effectively represents the peak coupling strength relevant for resonator optimization, even though the actual excitonic wavefunction is spatially extended. A more complete treatment would involve distributed sources~\cite{Carminati_2022} or susceptibility-based modeling, but this is beyond the scope of this work and not required for comparing the different resonator designs.

\begin{figure}
\includegraphics[width=0.49\textwidth]{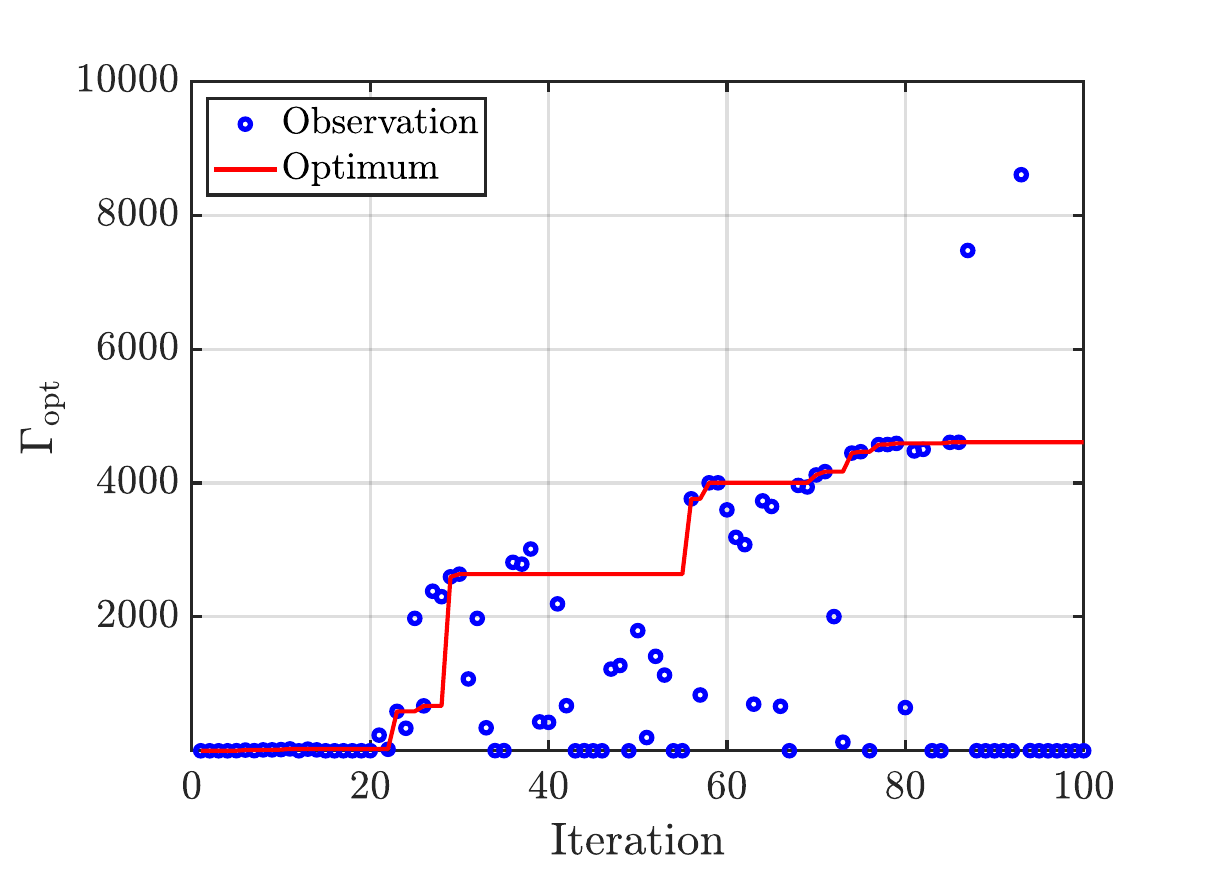}
\caption{\label{fig3}
Optimization of the nanobeam resonator. The optimization parameters are the lattice constant $a$, the width $w$ of the resonator, and the thickness $t$ of the individual $\mathrm{WS}_2$ layers. The target quantity $\Gamma_\mathrm{opt}$ is defined by Eq.~\eqref{gamma_opt} and it is related to the maximum Purcell enhancement of the resonator near the emission wavelength of the 2D exciton of the MoSe$_2$ layer. In each iteration step, an observation $\Gamma_\mathrm{opt}$ is calculated. These observations are marked with blue dots. The maximum value of the observations reached is marked by the red fragmented line. The blue dots above this line correspond to observations with $Q \geq 10^5$, i.e., outside the imposed $Q$-factor constraint.}
\end{figure}

In practice, the nanolaser could initially be operated under optical pumping for a proof-of-concept demonstration, whereas future implementations may employ electrical injection following established TMDC-based injection schemes~\cite{Chen_2025}. Such injection may introduce moderate additional loss and dephasing~\cite{Kekatpure_2008,Kumar_2014,Moody_2015,Mueller_2015,Dobusch_2017} but does not affect the fundamental resonator design principles presented here.

The nanobeam resonator is spatially discretized using the finite element method (FEM). We use the software package JCMsuite~\cite{Pomplum_NanoopticFEM_2007} for the numerical discretization and for the computation of the resonances to calculate the Purcell enhancement using Eq.~\eqref{eq:purcell_expansion}. Perfectly matched layers are applied for the realization of the open boundary conditions in the $x$, $y$ and $z$ directions. In the numerical implementations, the mirror symmetries in the $x$ and $y$ directions are exploited. The mesh for the numerical discretization is refined a-priori and the convergence of the numerical solutions is ensured by choosing a suitable degree $p$ of the polynomial ansatz functions of the FEM implementation. An increase of $p$ leads to a higher number of unknowns of the implementation, i.e., the accuracy and the computation time increase. The corresponding numerical settings can be found in the data publication~\cite{Binkowski_SourceCode_Nanobeam}. We note that the numerical precision stated in this work is related to the accuracy of the simulations and may not be achievable in experimental implementations.

\subsection{Optimization}
We apply a Bayesian optimization algorithm~\cite{Pelikan_1999,Schneider_Benchmark_2019} to optimize the Purcell enhancement of the nanobeam resonator. This global optimization approach is well suited for problems with computationally expensive objective functions, as is the case with the 3D implementation of the resonator under consideration. However, other optimization approaches can be used as well.

We choose the parameter ranges $140\,\mathrm{nm} \leq a \leq 340\,\mathrm{nm}$, $160\,\mathrm{nm} \leq w \leq 360\,\mathrm{nm}$, and $100\,\mathrm{nm} \leq t \leq 300\,\mathrm{nm}$. This leads to a minimum distance of $20\,\mathrm{nm}$ between the holes in the $x$ direction and to the outer air region in the $y$ direction. In each optimization step, we compute the $N=16$ resonances of the resonator which resonance wavelengths are closest to the emission wavelength of the 2D exciton of the MoSe$_2$ layer, given by $\lambda = 780\,\mathrm{nm}$. This means that Eq.~\eqref{eq:maxwell} without a source term is solved. Then, we calculate the Purcell enhancement $\Gamma(\lambda)$ using Eq.~\eqref{eq:purcell_expansion} based on these modes. The obtained resonance expansion $\sum_{n=1}^{16} \Gamma_n(\lambda)$ is then evaluated for a fine sampling in the wavelength range $760\,\mathrm{nm} \leq \lambda \leq 800\,\mathrm{nm}$.
The objective function of the optimization is the quantity
\begin{equation} \label{gamma_opt}
\Gamma_\mathrm{opt} = \max\limits_{\lambda} \left( f_\mathrm{p}(\lambda) \sum_{n=1}^{16} \Gamma_n(\lambda)\right)
\end{equation}
within this wavelength range. The shape function $f_\mathrm{p}(\lambda)$ is a parabola with the maximum at the target wavelength $\lambda = 780\,\mathrm{nm}$ and the zeros at the edges of the wavelength range. By applying this function, the optimization can steer the spectral position of the maximum Purcell enhancement towards the desired target wavelength while ensuring that the spectral peak remains within the selected wavelength range.

\begin{figure}
\includegraphics[width=0.49\textwidth]{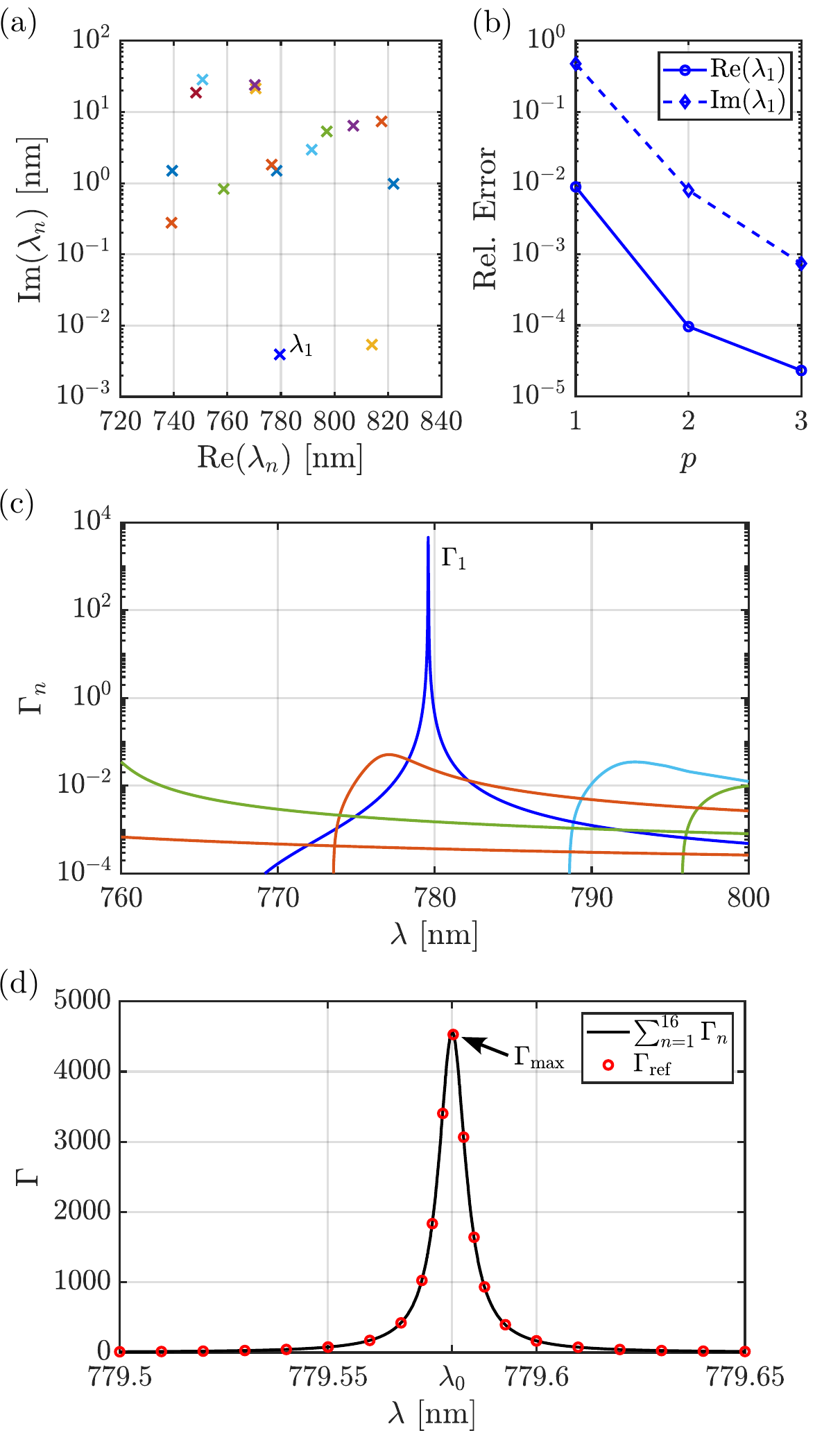}
\caption{\label{fig4}
Optimization results of the nanobeam resonator. (a)~Resonance wavelengths $\lambda_n$ of the optimized resonator. The resonance wavelength  $\lambda_1 =(779.580+ 0.0039i)\,\mathrm{nm}$ has the smallest imaginary part. (b)~Relative errors $|\mathrm{Re}(\lambda_1(p))-\mathrm{Re}(\lambda_{1,\mathrm{ref}})|/\mathrm{Re}(\lambda_{1,\mathrm{ref}})$ and $|\mathrm{Im}(\lambda_1(p))-\mathrm{Im}(\lambda_{1,\mathrm{ref}})|/\mathrm{Im}(\lambda_{1,\mathrm{ref}})$, where the reference solutions are computed using the degree $p=4$ of the polynomial ansatz functions of the FEM implementation. (c)~Modal contributions $\Gamma_n$ corresponding the resonance wavelengths, within the wavelength range from the optimization. The most significant contribution is $\Gamma_1$. (d)~Sum $\sum_{n=1}^{16} \Gamma_n$ over the $N=16$ modal contributions and reference solutions $\Gamma_\mathrm{ref}$ computed by solving scattering problems given by Eq.~\eqref{eq:maxwell}, shown in a sub-nanometer wavelength range. The maximum Purcell enhancement $\Gamma_\mathrm{max} = 4.563\times10^3$ and the spectral position $\lambda_0 =  779.580\,\mathrm{nm}$ of the maximum are marked.}
\end{figure}

\begin{figure*}
\includegraphics[width=0.98\textwidth]{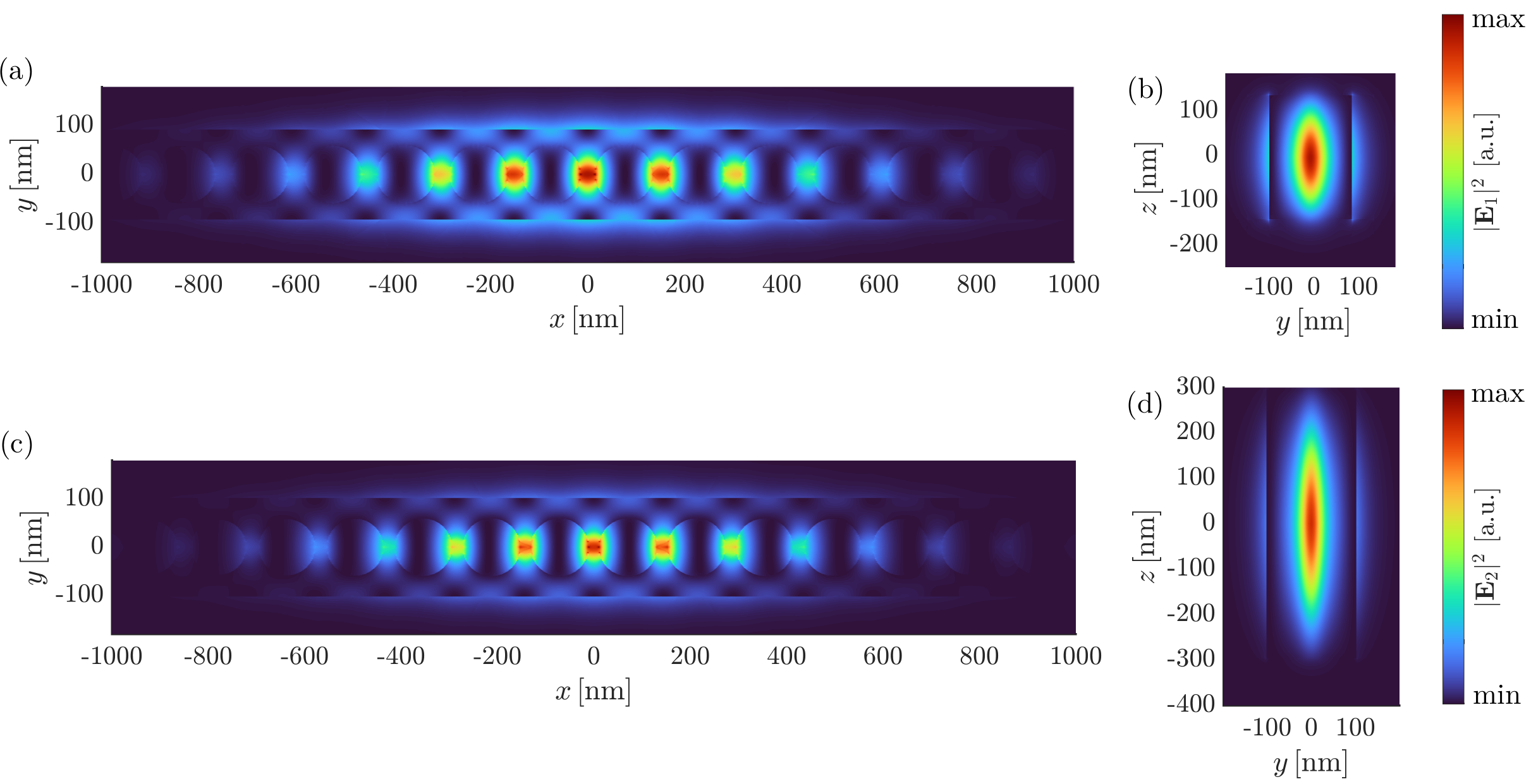}
\caption{\label{fig5}
Visualization of the electric field intensities of central detail sections of resonance modes.
The mode $\mathbf{E}_1$ corresponds to the optimization with a constraint on the $Q$-factor, which results in a Purcell enhancement of $4.6\times10^3$ and a $Q$-factor of $9.92\times10^4$. The mode $\mathbf{E}_2$ corresponds to the optimization without a constraint on the $Q$-factor, which results in a Purcell enhancement of $2.4\times 10^{4}$ and a $Q$-factor of $6.42\times 10^5$. (a)~Intensity $|\mathbf{E}_1|^2$ in the $xy$ plane. (b)~Intensity $|\mathbf{E}_1|^2$ in the $yz$ plane. (c)~Intensity $|\mathbf{E}_2|^2$ in the $xy$ plane. (d)~Intensity $|\mathbf{E}_2|^2$ in the $yz$ plane. All field intensities are shown in planes through the center of the system. The plots in (a) and (b) and the plots in (c) and (d) each share one colorbar. The scaling in (a) and (b) is not the same as in (c) and (d).}
\end{figure*}

We further impose a constraint on the $Q$-factor, requiring $Q < 10^5$. While a high $Q$-factor theoretically enhances the Purcell effect by increasing the light–matter interaction strength, an excessively high $Q$-factor can be problematic for real-world applications~\cite{Lodahl_2015}. Systems with a very sharp spectral peak of the Purcell enhancement are highly sensitive to fabrication imperfections and the underlying resonance is highly sensitive to wavelength mismatches with a realistic emitter, leading to a loss of spectral overlap and a reduced enhancement. Additionally, many realistic emitters have broader spectral linewidths that do not couple efficiently to resonances with too high $Q$-factors.

Figure~\ref{fig3} shows the target quantity $\Gamma_\mathrm{opt}$ over the iteration number of the optimization. A maximum number of $100$ iterations is selected.
At the $86$th iteration, with the parameter combination of $a = 149.68\,\mathrm{nm}$, $w = 184.98\,\mathrm{nm}$, and $t = 138.05\,\mathrm{nm}$, we obtain an optimum of $\Gamma_\mathrm{opt}= 4.6\times10^3$. Note that a higher Purcell enhancement can also be achieved if, e.g., more iterations are performed or the parameter ranges are changed. Such investigations are beyond the scope of this work.

\subsection{High Purcell enhancement}
Figure~\ref{fig4}(a) shows the resonance wavelengths $\lambda_n=2\pi c/\omega_n$, where $\omega_n$ are the resonance frequencies, of the optimized nanobeam resonator. The wavelength $\lambda_1 =(779.580+ 0.0039i)\,\mathrm{nm}$ has the smallest imaginary part. The corresponding $Q$-factor is given by $Q = 9.92\times10^4$. Figure~\ref{fig4}(b) shows the convergence of $\lambda_1$ with respect to the degree $p$ of the polynomial ansatz functions of the FEM implementation. For $p=2$, accuracies of $9.6\times10^{-5}$ and $7.9\times10^{-3}$ are obtained for the real and imaginary part, respectively. The degree $p=2$ is chosen for the optimization shown in Fig.~\ref{fig3} and $p=4$ is selected for computing the results in Fig.~\ref{fig4}. The optimization with a numerical implementation with fewer unknowns and the subsequent more accurate recalculation of the result reduces the total computational effort. 

Figure~\ref{fig4}(c) shows the single contributions $\Gamma_{n}$ within the wavelength range from the optimization. The significant contribution is given by $\Gamma_1$. The other contributions are negligible in the considered wavelength range. Figure~\ref{fig4}(d) shows $\sum_{n=1}^{16} \Gamma_n$ in a small wavelength range for the optimized nanobeam resonator. The maximum $\Gamma_\mathrm{max} = 4.563\times10^3$ occurs at the wavelength $\lambda_0 =  779.580\,\mathrm{nm}$. The spectral linewidth at half of the maximum Purcell enhancement is $0.008\,\mathrm{nm}$. A Purcell enhancement greater than $50$ is maintained for wavelength variations up to $0.037\,\mathrm{nm}$. We also show reference solutions $\Gamma_\mathrm{ref}$ which are computed by solving scattering problems given by Eq.~\eqref{eq:maxwell} using the subtraction field approach given by Eq.~\eqref{eq:Maxwell_correction}. It can be observed that $\sum_{n=1}^{16} \Gamma_n$ coincides with $\Gamma_\mathrm{ref}$. This emphasizes that considering $N=16$ resonances for the computation of the total Purcell enhancement $\Gamma$ using Eq.~\eqref{eq:purcell_expansion} is sufficient.
The $Q$-factors of the resonance frequencies in the wavelength range $760\,\mathrm{nm} \leq \mathrm{Re}(\lambda_n) \leq 800\,\mathrm{nm}$ are below 300, and the corresponding electric field overlaps of the resonance modes with the monolayer are also substantially small. Consequently, the high-$Q$ mode corresponding to $\lambda_1$ remains the dominant channel for cavity-enhanced emission, while the nearby leaky modes contribute only minor radiative losses and have a negligible impact on the overall Purcell enhancement.
Note that, since the high $Q$-factor leads to a sharp spectral peak of the Purcell enhancement, fine sampling is required in the wavelength range under consideration. This is possible because the evaluation of the resonance expansion in Eq.~\eqref{eq:purcell_expansion} causes a negligible computational effort for the different wavelengths.

Figures~\ref{fig5}(a) and ~\ref{fig5}(b) show the intensities of the electric field of the resonance mode corresponding to the most significant resonance wavelength $\lambda_1$ in the $xy$ and $yz$ plane, respectively. It can be observed that the maximum of the intensity is located at the center of the resonator, where the location of the dipole emitter is assumed. The intensity of the electric field decreases in the $x$ direction towards the outer regions of the nanobeam resonator.

Note that higher Purcell enhancements than those presented in this work can be achieved when the $Q$-factor is not limited. To demonstrate this, we perform an additional optimization without imposing a constraint on the $Q$-factor, which serves as an example for an extreme high achievable Purcell enhancement in our system. This optimization yields a maximum Purcell enhancement of $2.4\times 10^{4}$, where the dominant resonance has a wavelength of $(780.290 + 0.0006i)\,\mathrm{nm}$ and a $Q$-factor of $6.42\times 10^5$. The parameter ranges and the target quantity are the same as for the other optimization. The parameter combination of the optimum is given by $a = 141.47\,\mathrm{nm}$, $w = 204.37\,\mathrm{nm}$, and $t = 294.27\,\mathrm{nm}$. Details on this additional optimization can be found in the data publication~\cite{Binkowski_SourceCode_Nanobeam}. Such an unconstrained optimization example is even more affected by the fact that realistic emitters with broader spectral linewidths may not couple efficiently to resonances with excessively high $Q$-factors~\cite{Bjoerk_1993}. However, if both designs were realized experimentally, the design with the higher $Q$-factor could still benefit more from it than the one obtained from the constrained optimization, even though in both cases the theoretically predicted $Q$-factor cannot be fully exploited. Figures~\ref{fig5}(c) and~\ref{fig5}(d) show the electric field intensities of the corresponding resonance mode $\mathbf{E}_2$. It can be observed that the intensity decreases more rapidly towards the outer regions of the system compared to the intensity of the resonance mode $\mathbf{E}_1$ from the other optimization, which exhibits a lower $Q$-factor.

\subsection{Competing nanolaser designs}
For a comparison with recently reported laser geometries, we consider the Purcell factor $F_n$ associated with the dominant resonance mode $\mathbf{E}_n(\mathbf{r})$, which has the resonance wavelength $\lambda_n$ and the quality factor $Q$. An alternative to computing $F_n$ from Eq.~\eqref{eq:purcell_expansion}, where $F_n = \Gamma_n(\mathrm{Re}(\lambda_n))$, is the commonly used expression
\begin{align}
    F_n = \frac{3}{4\pi^2}\left( \frac{\mathrm{Re}(\lambda_n)}{n} \right)^3 Q \mathrm{Re}\left( \frac{1}{V_n}\right), \label{eq:purcell_QV}
\end{align}
where
\begin{align}
    V_n = \frac{1}{2\epsilon_0 n^2\left( \mathbf{E}_n(\mathbf{r}^\prime)\cdot \hat{\mathbf{j}}\right)^2}  \label{eq:mode_volume}
\end{align}
is the complex-valued mode volume~\cite{Sauvan_QNMexpansionPurcell_2013,Lalanne_QNMReview_2018} with the normalized dipole strength vector $\hat{\mathbf{j}} = \mathbf{j}/|\mathbf{j}|$ and
$n$ is the refractive index at the dipole emitter position $\mathbf{r}^\prime$. For this expression, the resonance modes are normalized such that $\int_\Omega \left[ {\mathbf{E}_n}(\mathbf{r})\cdot \frac{\partial \omega \epsilon(\mathbf{r},\omega)}{\partial \omega} {\mathbf{E}_n}(\mathbf{r}) - \mu_0 {\mathbf{H}_n}(\mathbf{r})\cdot {\mathbf{H}_n}(\mathbf{r}) \right] dV = 1$, where $\mathbf{H}_n$ is the magnetic field corresponding to the resonance mode and $\Omega$ is the computational domain. Note that, in the literature, the computation of the mode volume $V_n$ can differ from Eq.~\eqref{eq:mode_volume}. For instance, some works apply a different normalization of the resonance modes or assume an optimal emitter position and orientation, rather than evaluating the mode volume at the actual dipole position and orientation. Nevertheless, for a qualitative comparison, we contrast the Purcell factor $F_n$ and the mode volume $V_n$ of our all-TMDC nanobeam resonator with those from other works, without recomputing the published results according to Eq.~\eqref{eq:mode_volume}. We further note that very high theoretically predicted Purcell factors may not be achieved when fabrication tolerances are taken into account~\cite{Plock_2024}.

We first consider the monolayer excitonic laser demonstrated in Ref.~\cite{Ye_2015}. In that work, a WSe$_2$ monolayer is coupled evanescently to a Si$_3$N$_4$ microdisk resonator with a quality factor of $Q \approx 2.6\times10^3$ and a mode volume of $V_n \approx 5 (\mathrm{Re}(\lambda_n)/n)^3$, leading to an estimated Purcell factor of $F_n \approx 40$. The moderate $F_n$ can be attributed to the evanescent coupling of the monolayer positioned on top of the microdisk. For a representative example of a conventional dielectric nanobeam resonator, we refer to the design reported in Ref.~\cite{Quan_2011}. The proposed silicon photonic crystal nanobeam resonator achieves a simulated quality factor of $Q \approx 5\times10^9$ with a mode volume of $V_n \approx 0.9(\mathrm{Re}(\lambda_n)/n)^3$, yielding a Purcell factor of $F_n \approx 4.2\times10^8$ under optimal dipole alignment.
The $Q$-factor-constrained optimization of our all-TMDC nanobeam resonator design yields a resonance with $Q = 9.92\times10^4$, a mode volume of $\mathrm{Re}(V_1) = 1.65(\mathrm{Re}(\lambda_1)/n)^3$, and a numerically optimized Purcell factor of $F_1 = 4.6\times10^3$. Note that evaluating 
$\Gamma_1(\mathrm{Re}(\lambda_1))$ from Eq.~\eqref{eq:purcell_expansion} and $F_1$ from Eq.~\eqref{eq:purcell_QV} yields identical results.

While $F_1 = 4.6\times10^3$ remains below the ultimate theoretical limit predicted for ultra-high-$Q$ dielectric nanobeam resonators, it already exceeds that of the TMDC microdisk laser by roughly two orders of magnitude. Moreover, in our optimized design, the monolayer emitter coincides with the field maximum, in contrast to the evanescent coupling geometry from Ref.~\cite{Ye_2015}.

\section{Conclusion}
An all-TMDC nanobeam resonator with an active monolayer that supports high Purcell enhancement was proposed and investigated numerically. The system is designed to function as a nanolaser. A numerical optimization resulted in a resonator design supporting a resonance mode with a localization of the electric field at the center of the system and with a corresponding resonance frequency close to the emission frequency of the active layer. The resonance exhibits a $Q$-factor of $9.92\times 10^{4}$ and enables a high Purcell enhancement of~$4.6\times10^3$.

For modeling and optimizing the Purcell enhancement of dipole emitters in atomically thin layers, we proposed a framework based on the calculation of resonance modes and their resonance frequencies, i.e., on the solution of Maxwell's equation without a source term. The resonance modes and resonance frequencies are used for a resonance expansion of the Purcell enhancement. In this way, the sharp spectral peak of the Purcell enhancement, which occurs due to the high-$Q$ resonance, can be efficiently resolved and optimized. A constraint on the $Q$-factor was imposed to obtain a not too narrow spectral linewidth for realistic applications. In order to demonstrate the theoretical limits of our system, an additional optimization without constraining the $Q$-factor was performed. This resulted in a Purcell enhancement of $2.4\times 10^{4}$, where the underlying resonance has a $Q$-factor of $6.42\times 10^5$.

The proposed nanobeam resonator with technologically implementable dimensions can be a starting point for experimental realizations. We expect that the optimization framework based on resonance expansion will be useful to propose and investigate further resonator designs with high Purcell enhancements. Although we demonstrated design concepts in the context of a nanobeam resonator, the framework presented is general and can be adapted to a broad range of resonator types. Photonic crystal cavities or microring resonators offer similar or complementary figures of merit and could enable alternative coupling configurations or footprint optimizations~\cite{Sriram_2020,Munkhbat_2023}. Furthermore, the high refractive index of TMDCs makes them well suited as a platform for diverse resonator geometries, paving the way toward monolithic, reconfigurable, and ultra-compact 2D material photonic circuits~\cite{Kim_2023}.

\section*{Data availability} 
Source code and simulation results for the numerical experiments for this work can be found in the open access data publication~\cite{Binkowski_SourceCode_Nanobeam}.

\section*{Acknowledgments}
We acknowledge funding 
by the Deutsche Forschungsgemeinschaft (DFG, German Research Foundation) under Germany's Excellence Strategy - The Berlin Mathematics Research Center MATH+ (EXC-2046/1, project ID: 390685689), 
by the German Federal Ministry of Research, Technology and Space (BMFTR, Forschungscampus MODAL, project 05M20ZBM),
by the Senate of Berlin, within the Program for the Promotion of Research, Innovation and Technology (ProFIT) co-financed by the European Regional Development Fund (ERDF, application no.~0206824, SQALE),
by the European Research Council (ERC-StG ``TuneTMD'', grant no.~101076437), 
and by the Villum Foundation (grant no.~VIL53033).

\end{document}